\documentclass{PoS}
\usepackage{graphicx}
\usepackage{amsmath}

\title{Analysis on RXTE, INTEGRAL and ROTSE IIId observations of the X-ray Pulsar 4U 1907+09}

\ShortTitle{Analysis on Observations of the X-ray Pulsar 4U 1907+09}

\author{\speaker{{\c{S}}eyda {\c{S}}ahiner} \\
        Middle East Technical University\\
        E-mail: \email{seyda@astroa.physics.metu.edu.tr}}

\author{S{\i}tk{\i} {\c{C}}a{\u{g}}da{\c{s}} {\.{I}}nam \\
        Ba{\c{s}}kent University \\
        E-mail: \email{inam@baskent.edu.tr}}
\author{Altan Baykal \\
		Middle East Technical University\\
		E-mail: \email{altan@astroa.physics.metu.edu.tr}}
\author{\"{U}mit K{\i}z{\i}lo\u{g}lu \\
		Middle East Technical University\\
		E-mail: \email{umk@astroa.physics.metu.edu.tr}}

\abstract{In this paper we present our recent timing and spectral analysis of the X-ray pulsar 4U 1907+09. Our X-ray data 
consist of an extended set of RXTE \& INTEGRAL observations that were analyzed before ({\c{S}}ahiner et al. 2012). From the X-ray observations we extend the pulse period history of the source 
and obtain a revised orbital distribution of the X-ray dips. Using ROTSE IIId optical observations, we present the long term optical 
light curve of the source to have an understanding of long
term optical behaviour.}

\FullConference{ An INTEGRAL view of the high-energy sky (the first 10 years) - 9th INTEGRAL Workshop and celebration of the 10th anniversary of the launch\\
                 15-19 October 2012\\
                 Bibliotheque Nationale de France, Paris, France}

\begin{document}

\section{Introduction}
4U 1907+09 contains an accretion powered X-ray pulsar and a blue supergiant companion star. The eccentricity and orbital period of the system are $\sim$ 0.28 and 
$\sim$8.3753 days respectively (in 't Zand et al. 1998). As 4U 1907+09 shows two phase locked flares per orbit separated 
by $\simeq0.5$ in orbital phase, the companion star was thought to be of B\emph{e} type and the compact object passing through 
a circumstellar disk of the equatorial plane of this companion (Iye 1986, Cook \& Page 1987). However optical 
(Cox et al. 2005) and infrared (Nespoli et al. 2008) observations showed that the companion could be classifed as an 
O8 - O9 Ia supergiant with a minimum distance of $\sim$5 kpc. 

The spin period of 4U 1907+09 was first measured as $\sim$437.5 s using \emph{Tenma} observations (Makishima et al. 1984). 
After the pulsar had been observed to be steadily spinning down until 1998 with an average rate of 
$\dot{\nu} = -3.54 \times 10^{-14}$ Hz s$^{-1}$, \emph{RXTE} observations showed that the spin rate got lower by a factor of 
$\sim$0.60 (Baykal et al. 2006). From \emph{INTEGRAL} observations, it was found that the spin period had reached a 
maximum of $\sim$441.3 s then, a torque reversal occurred and the source began to spin up with a rate of 
$2.58 \times 10^{-14}$ Hz s$^{-1}$ after 2004 May (Fritz et al. 2006). Recent measurements
(see {\c{S}}ahiner et al. (2012) and references therein) have indicated that 4U 1907+09 has returned to a 
new spin-down phase which is similar to the previous spin-down with a rate of $-3.59 \times 10^{-14}$ Hz s$^{-1}$. 

4U 1907+09 is a source showing irregular flaring and dipping episodes. in 't Zand et al. (1997) reported that $\sim$20\% 
of observations show a dip state with no detectable pulsed emission.  The presence of dipping states are thought to be related 
with the cessation of the accretion from an inhomogeneous wind of the companion star. 

In this paper we extend previous timing and spectral analysis of \emph{RXTE} \& \emph{INTEGRAL} observations
(see {\c{S}}ahiner et al. (2012) and references therein).  Using \emph{ROTSE IIId} optical 
observations, we also present the long term optical light curve of the source. In the next section, we present observations. 
In Section 3, we discuss our analysis and conclude.

\section{Observations}

\subsection{RXTE and INTEGRAL}

In this paper, we present results of the analysis of 118 pointed \emph{RXTE}-PCA between
2007 June and 2011 December each with an exposure of $\sim$2 ks. 98 of these observations was previously analyzed (see 
{\c{S}}ahiner et al. (2012) and references therein). Along with the \emph{RXTE}-PCA data, 
we also use the previous analysis results of \emph{INTEGRAL} observations ({\c{S}}ahiner et al. 2012) between 2005 October 
and 2007 November.

For the recent 20 \emph{RXTE}-PCA observations, we use the same data selection criteria as before by {\c{S}}ahiner et al. 
(2012):  For spectral analysis, we only use data obtained from PCU 2 in order to avoid probable problems due to calibration 
differences between the detectors, but we do not make any PCU selection for the timing analysis since the loss of propane 
layers of PCU 0 and PCU 1 do not affect high resolution timing.

The standard software tools of \verb"HEASOFT v.6.10" are used for the analysis of PCA data. Background spectra and light 
curves are generated by the latest PCA background estimator models supplied by the \emph{RXTE} Guest Observer Facility (GOF)
, \verb"Epoch 5C".

\subsection{ROTSE}

The optical data were obtained with Robotic Optical Transient Experiment \footnote{http://www.rotse.net}
(ROTSE IIId) located at  Bak{\i}rl{\i}tepe, Antalya, Turkey\footnote{http://www.tug.tubitak.gov.tr}. ROTSE IIId operates 
without filters thus have a passband peaking around 550nm (Akerlof et al. 2003). Optical analysis is  performed for the 
daily observations of the source between 2004 July and 2010 September. Approximately 1700 CCD frames with 20 second exposure 
time are used. All images are automatically dark and flat field corrected by a data extraction pipeline of ROTSE IIId 
(Akerlof et al. 2003). Instrumental magnitudes are obtained using aperture photometry on the observed CCD frames. ROTSE 
magnitudes are calibrated by comparing all the field stars against the USNO A2.0 R-band catalog.  Barycentric corrections 
are applied to the times of each  observation by using JPL DE200 ephemerides. Details on the reduction pipeline of data 
were described in K{\i}z{\i}lo\u{g}lu et al. (2005).

\begin{figure*}
  \center{\includegraphics[width=12cm, angle=0]{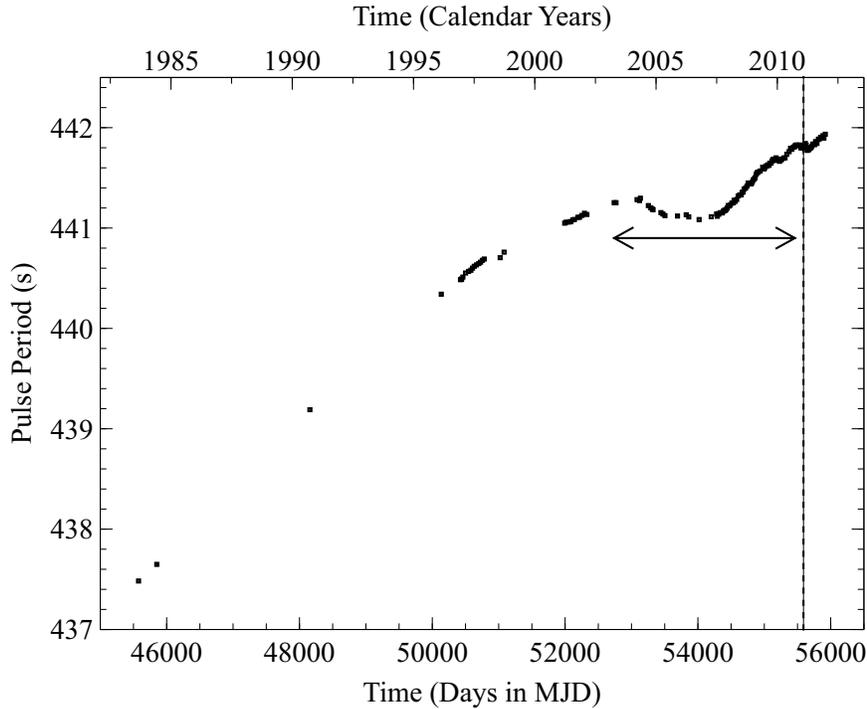}} 
  \caption{Pulse period history of 4U 1907+09. The measurements with \emph{RXTE} and \emph{INTEGRAL} in this work and 
  previous studies (see {\c{S}}ahiner et al. (2012) and references therein) are included. The pulse period values calculated in this study lie on the right of the vertical dashed line. Horizontal line with arrows indicates time span of the ROTSE observations of the source.}
  \label{history}
\end{figure*}

\section{Analysis}

\begin{figure*}
  \center{\includegraphics[width=7cm, angle=270]{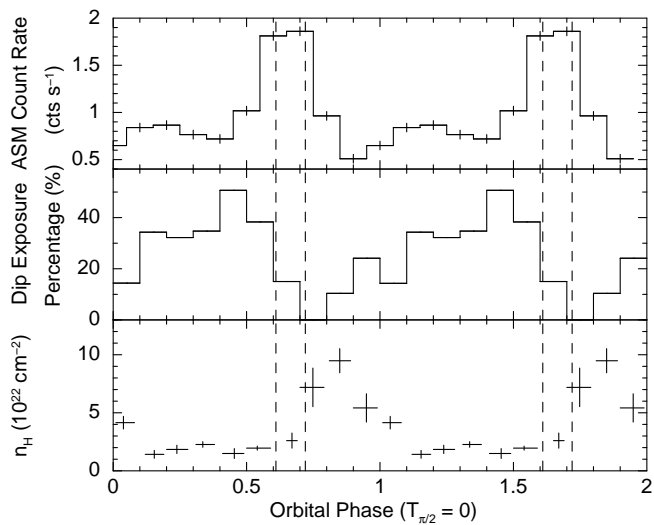}} 
  \caption{ASM light curve folded at the orbital period (top panel), the percentage of dip exposure times to the total 
  exposure through the binary orbit (middle panel) and variation of n$_H$ from non-dipping observations (bottom panel; see also 
  Figure 5 of {\c{S}}ahiner et al. 2012)). The vertical dashed lines correspond to the time of periastron passage within
  1$\sigma$. This figure is a revised version of Figure 8 by {\c{S}}ahiner et al. (2012) including new dipping states.}
  \label{dipping}
\end{figure*}

\begin{figure*}
  \center{\includegraphics[width=7cm, angle=270]{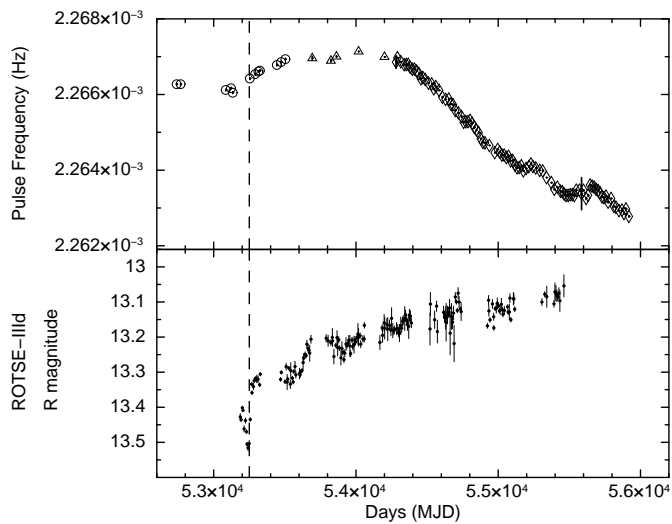}} 
  \caption{Long term spin frequency history (top panel) and 7-day avaraged long term \emph{ROTSE IIId} light curve 
  (bottom panel) of 4U 1907+09. Vertical dashed line indicates the time around when the the optical luminosity of the 
  source drops and this drop is restored quickly.}
  \label{rotse}
\end{figure*}

For timing analysis, 1 s binned background and solar barycenter corrected \emph{RXTE}-PCA light curve of the source is used 
to extend the analysis explained by {\c{S}}ahiner et al. (2012). These light curves are also corrected to account for the 
binary motion of 4U 1907+09 using the binary orbital parameters (in 't Zand et al. 1998). Dips from the light curves are 
also eliminated. 

Pulse periods for 4U 1907+09 are found by folding the time series on statistically 
independent trial periods (Leahy et al. 1983) and template pulses are generated from these observations by folding 
the data on the period giving maximum $\chi^2$. Each pulse profile which is represented by 
its Fourier harmonics (Deeter \& Boynton 1985) contains 20 phase bins. We obtain pulse arrival times from the 
cross-correlation between 
template and pulse profiles obtained in each $\sim$2 ks observation. 

To obtain the pulse period of the source, we continue to fit newer pulse arrival pairs to a linear model as 
{\c{S}}ahiner et al. (2012) did before for the previous observations and estimate the pulse period values at the mid-time of the 
observations. The  pulse period history of the source is presented in Figure \ref{history} including results of analysis 
of this paper and the other \emph{RXTE} and \emph{INTEGRAL} papers (see {\c{S}}ahiner et al. (2012) and references therein).

As an extension to the dipping state analysis of {\c{S}}ahiner et al. (2012), we add 12 dipping states from the most 
recent \emph{RXTE}-PCA observations (proposal ID: P96366) and revise Figure 8 of that study (see Figure \ref{dipping}). 

Finally to investigate the long term optical variability of the source we construct the long term optical light curve of the 
source using \emph{ROTSE IIId} observations (see Figure \ref{rotse}). Time span of these observations is indicated in Figure \ref{history}. In Figure \ref{rotse}, we also include long term 
pulse frequency history of the source to see whether there is any correlation between the spin rate and the optical light 
curve of the source or not. 

\section{Conclusion}

From the timing analysis, we find that long term spin-down trend of the source continues (see Figure \ref{history}).   

The percentage of exposure time spent in dipping is reduced from 28 to 24 per cent since the recently analyzed set of observations includes less dipping 
episodes (see Figure \ref{dipping}). The orbital dependence of dip occurrence stated by {\c{S}}ahiner et al. (2012) is preserved, i.e. dipping states are frequently observed between the orbital phases through 0.1 to 0.6 and they are not observed between phases 0.7 and 0.8. 

Although a direct spectral study of the dips with \emph{RXTE} is not possible because of the diffuse galactic emission 
background, comparing the orbital dependence of dipping with the spectral results of non-dipping observations 
(see the bottom panel of Figure \ref{dipping}) an anti-correlation is recognized, such that the absence of dips matches with 
the orbital phases when n$_H$ is maximum just after the periastron passage and frequently dipping phases match with the times 
of minimum n$_H$. This result implies that the dipping states are not due to increased absorption, in fact it supports the 
idea that the dips arise from the clumpy nature of the companion wind. As the pulsar passes through low-density regions of 
the wind, the accretion rate decreases, the Alfv\'{e}n radius increases and accretion stops when the source enters the 
propeller regime. The frequent dipping states observed after the apastron until the periastron, are candidate episodes for 
transition to a temporary propeller state due to accretion from the clumpy wind.

A recent study on the dipping episodes of 4U 1907+09 observed by \emph{Suzaku} (Doroshenko et al. 2012) reveals that the pulsations are sustained 
during these episodes. This fact implies that the accretion does not stop completely, while a change in the accretion regime might be responsible 
for the decrease in the source flux. These authors also put a different approach forward by considering
transient flaring episodes rather than transient dipping episodes and explained the flare-like events between 
consecutive dips by the gated accretion scenario proposed to explain the flares in SFXTs (Bozzo et al. 2008). The timescales and orbital phase 
dependence of flaring activity in 4U 1907+09 are similar to that of SFXTs, however the dipping states are shorter and brighter when compared to 
quiescence emission from SXFTs. The distinguishing property responsible for the difference is the compactness of the binary system, that is the 
compactness of 4U 1907+09 compared to SFXTs, causes relatively bright dipping episodes rather than deep quiescent episodes. Doroshenko et al. (2012) 
regards 4U 1907+09 as an intermediate system between the SFXTs and the persistent systems suggesting that it might be a missing link between the 
classes.

From the optical observations, we find that the optical luminosity of the source drops and this drop is restored quickly around 
MJD $\sim 53250$ when the source switches from an almost constant spin period phase to a spin-up phase (see Figure \ref{rotse}). Moreover, besides 
this short term variation, it is found that the optical luminosity of the source steadily increases and this trend does not 
alter whether the source spins up or down.

\section*{Acknowledgment}

We acknowledge support from T\"{U}B\.{I}TAK, the Scientific and Technological Research Council of Turkey through the 
research project TBAG 109T748.


\begin{thebibliography}{99}
\bibitem{} Akerlof C. W., Kehoe R. L., McKay T. A., et al. 2003, PASP, 115, 132
\bibitem{} Baykal A., \.{I}nam S. \c{C}., Beklen E., 2006, MNRAS, 369, 1760
\bibitem{} Bozzo E., Falanga M., \& Stella L. 2008, ApJ, 683, 1031
\bibitem{} Cook M. C., Page C. G., 1987, MNRAS, 225, 381
\bibitem{} Cox N. L. J., Kaper L., Mokiem M. R., 2005, A\&A, 436, 661
\bibitem{} Deeter J. E., Boynton P. E., 1985, in Hayakawa S. and Nagase F., Proc. Inuyama Workshop: Timing Studies of X-Ray 
Sources, p.29, Nagoya Univ., Nagoya
\bibitem{} Doroshenko V., Santangelo A., Ducci L. \& Klochkov D., 2012, A\&A, 548, 19
\bibitem{} Fritz S., Kreykenbohm I., Wilms J. et al., 2006, A\&A, 458, 885
\bibitem{} in 't Zand J. J. M., Strohmayer T. E., Baykal A., 1997, ApJ, 479, L47
\bibitem{} in 't Zand J. J. M., Baykal A., Strohmayer T. E., 1998, ApJ, 496, 386
\bibitem{} Iye M., 1986, PASJ, 38, 463
\bibitem{} Leahy D. A., Darbro W., Elsner R. F., Weisskopf M. C., Kahn S., Sutherland P. G., Grindlay J. E., 1983, ApJ, 266, 160
\bibitem{} Makishima K., Kawai N., Koyama K., Shibazaki N., 1984, PASJ, 36, 679
\bibitem{} K{\i}z{\i}lo\u{g}lu U., K{\i}z{\i}lo\u{g}lu N., \& Baykal A. 2005, AJ, 130, 2766
\bibitem{} Nespoli E., Fabregat J., Mennickent R. E., 2008, A\&A, 486, 911
\bibitem{} \c{S}ahiner \c{S}., \.{I}nam S.\c{C}., Baykal A., 2012, MNRAS, 421, 2079 

\end{thebibliography}
\end{document}